\documentclass{webofc}
\usepackage[varg]{txfonts}   

\usepackage{color} 

\newcommand{\HII}{H\,{\sc ii}}
   
\newcommand{\OI}{O\,{\sc i}}

\newcommand{\CII}{C\,{\sc ii}}

\begin{document}
\title{Multi-line Observations, Models, and
Data Needed to \mbox{Understand} the Nature of UV-irradiated Interstellar Matter }
%
%

\author{\firstname{Javier R.} \lastname{Goicoechea}\inst{1}\fnsep\thanks{\email{javier.r.goicoechea@csic.es}} \and
        \firstname{Sara} \lastname{Cuadrado}\inst{1}
        \and
        \firstname{Franck} \lastname{Le Petit}\inst{2}}

\institute{Instituto de F\'{\i}sica Fundamental (IFF),
     CSIC. Calle Serrano 121-123, 28006, Madrid, Spain. 
\and
LERMA, Observatoire de Paris, PSL University, CNRS, Sorbonne Universit\'e, 92190 Meudon, France.}

\abstract{%
Far-ultraviolet photons from OB-type massive stars regulate the \mbox{heating}, ionization, and chemistry of much of the neutral interstellar gas in star-forming galaxies. The interaction of  FUV radiation and interstellar \mbox{matter} takes place in
environments broadly known as \textit{photodissociation regions} (PDRs).  PDR line diagnostics are the smoking gun of the radiative feedback from \mbox{massive} stars. Improving our understanding of 
stellar feedback in the ISM \mbox{requires} quantifying the energy budget, gas dynamics, and chemical composition of PDR environments. This goal demands astronomical instrumentation able to deliver multi-line 
spectroscopic images of the ISM (of the Milky Way and nearby galaxies). 
It also requires interdisciplinary  collaborations to \mbox{obtain} 
the rate coefficients and cross sections of the many
microphysical processes that occur in the ISM and that are included in 
models such as the  \textit{Meudon} PDR code.

}
\maketitle
\section{Introduction}
\label{intro}

Far-UV  (FUV: $E$\,$<$\,13.6\,eV) photons emitted by   massive O- and B-stars govern, or at least greatly influence, the heating, ionization, and chemistry of the neutral 
interstellar gas: \mbox{everywhere} hydrogen atoms are in predominantly  neutral form.
The interaction of stellar FUV radiation and interstellar baryonic matter 
(atoms, molecules, and dust grains)
takes place in so-called \mbox{\textit{photodissociation regions}}
(PDRs;\,\cite{1999RvMP...71..173H}). 
This interaction occurs for different doses of FUV \mbox{radiation} ($G_0$) and
at  very different spatial scales: from the illuminated rims of star-forming clouds  to kpc scales in starburst galaxies.
The  emission from PDRs reflects the radiative feedback from massive stars,
a collection of processes that
establish the phases and pressures of the ISM,
and also regulate  star formation and molecular cloud destruction  \cite{2020A&A...639A...2P}.

The physical state of the interstellar gas depends on a plethora of detailed \mbox{microphysical} processes that determine its chemical composition and how the gas is heated and cooled. Therefore, in addition to understanding the (macro) astrophysical processes driving the \mbox{dynamics} and \mbox{evolution} of the ISM, it is also mandatory to understand the subtle \mbox{microprocesses} that form, destroy, and excite its basic constituents.
 Accompanied by new \mbox{developments} in astronomical \mbox{instrumentation}, last years 
on  PDR research have  \mbox{reinforced} closer collaborations with \mbox{molecular} physicists and laboratory experimentalists able to determine the rate coefficients and cross sections of the relevant microprocesses.

Bright molecular cloud rims, such as the iconic Orion Bar or the Horsehead,  have always been excellent laboratories to study  the molecular content of FUV-irradiated interstellar gas and  the wealth of physical and chemical processes that occur at microscopic (molecular) level.
 The  number of molecules, atoms, and ions  observed in these dense PDRs
    are
 not so easily detected toward fainter environments also influenced by stellar FUV radiation (e.g,  the surface layers of protoplanetary disks  and  distant star-forming galaxies).
 
In this invited paper we summarize some of the latest developments in observations and
modeling of PDRs. Due to the lack of space, we mostly limit to our own recent work.
\mbox{Readers} are referred to Wolfire \textit{et al.} \cite{2022arXiv220205867W} for a modern review on PDRs. Here we strengthen the \mbox{interdisciplinary} aspects of PDR research, and we emphasize the need of precise atomic and molecular data to correctly interpret  future observations
of the ISM in galaxies.

\begin{figure}[t]
\centering
\includegraphics[width=11.3cm,clip]{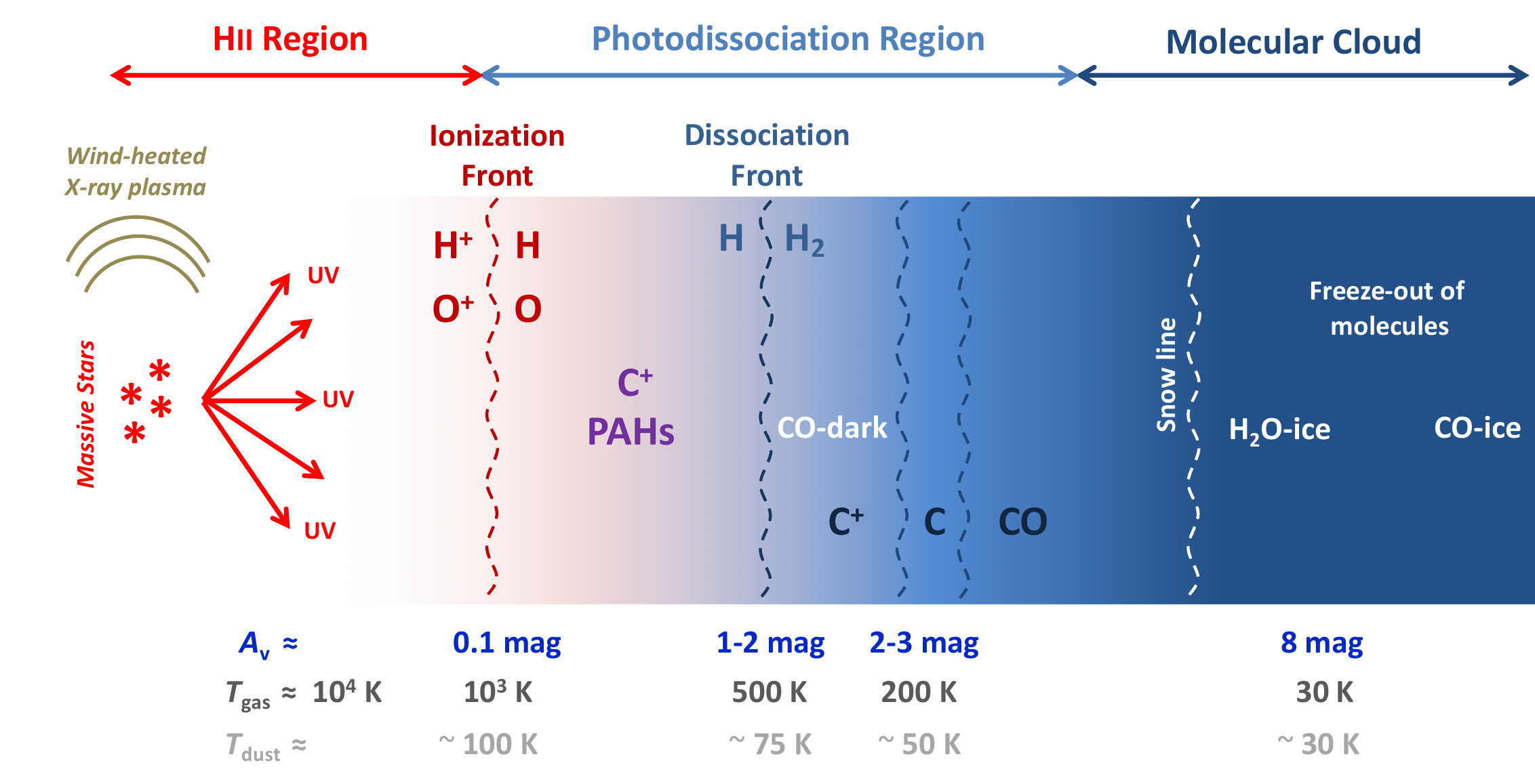}
\vspace{-0.4cm}
\caption{Structure of a strongly irradiated molecular cloud near massive stars (e.g., the Orion Bar).}
\label{fig-1}       
\end{figure}

\section{PDRs everywhere}

PDRs  host the critical conversion from atomic to molecular gas (the \mbox{H\,/\,H$_2$} and \mbox{C$^{+}$/\,C\,/\,CO} transition zones,  Fig.~1). This stratification occurs as the column density of gas and dust increases deeper inside molecular clouds, and as the  flux of FUV photons is attenuated by dust 
\mbox{(for high $G_0$\,/\,$n_{\rm H}$ ratios)} and  by H$_2$ absorption lines
(for low $G_0$\,/\,$n_{\rm H}$ ratios) \cite{2014ApJ...790...10S}. The penetration of FUV radiation
 depends on metallicity and grain properties. FUV photons reach larger depths in lower metallicity environments (lower dust-to-mass ratios)
 and in regions affected by grain growth (flatter extinction curves). Both effects
 shift the C$^+$/\,C\,/\,CO transition  to higher $A_V$ and leave a larger mass fraction
 of  H$_2$ gas that is \mbox{``CO-dark''} (or too faint). 

PDRs emit most of the IR radiation arising from the ISM of star-forming galaxies:
\mbox{FUV-pumped IR} bands from polycyclic aromatic hydrocarbons (PAHs), H$_2$ ro-vibrational, 
[\CII]158\,$\mu$m and [\OI]63,\,145\,$\mu$m fine-structure, and mid-$J$ CO lines,
as well as warm dust \mbox{continuum}  -- grains are heated by FUV photons and 
reemit FIR contimuum. Photoelectrons ejected from small grains and PAHs heat the gas. In  dense PDRs, collisional de-excitation of FUV-pumped H$_{2}$ is an important gas heating mechanism too.  
The presence of vibrationally excited \mbox{H$_{2}$($v$\,$\geq$\,1)} overcomes the endoergicity
 of the initiating reactions \mbox{H$_2$\,+\,(C$^+$, S$^+$, O, N, ...)} \cite{2010ApJ...713..662A} and leads  to the formation
 of CH$^+$, SH$^+$, OH, and NH \cite{2022arXiv220610441G}.

In their most general definition -- \textit{neutral gas regulated by FUV radiation} --, PDRs \mbox{represent} the  dominant fraction of the neutral atomic and molecular gas in the ISM of star-forming galaxies \cite{1999RvMP...71..173H,2022arXiv220205867W}. Indeed, except for the cold and dense molecular cores associated with the first stages of star formation, most of the ISM (in volume and mass) is effectively at $A_V$\,$<$\,8\,mag. Hence, permeated by stellar FUV photons.

\begin{figure}[t]
\centering
\includegraphics[width=12.5cm,clip]{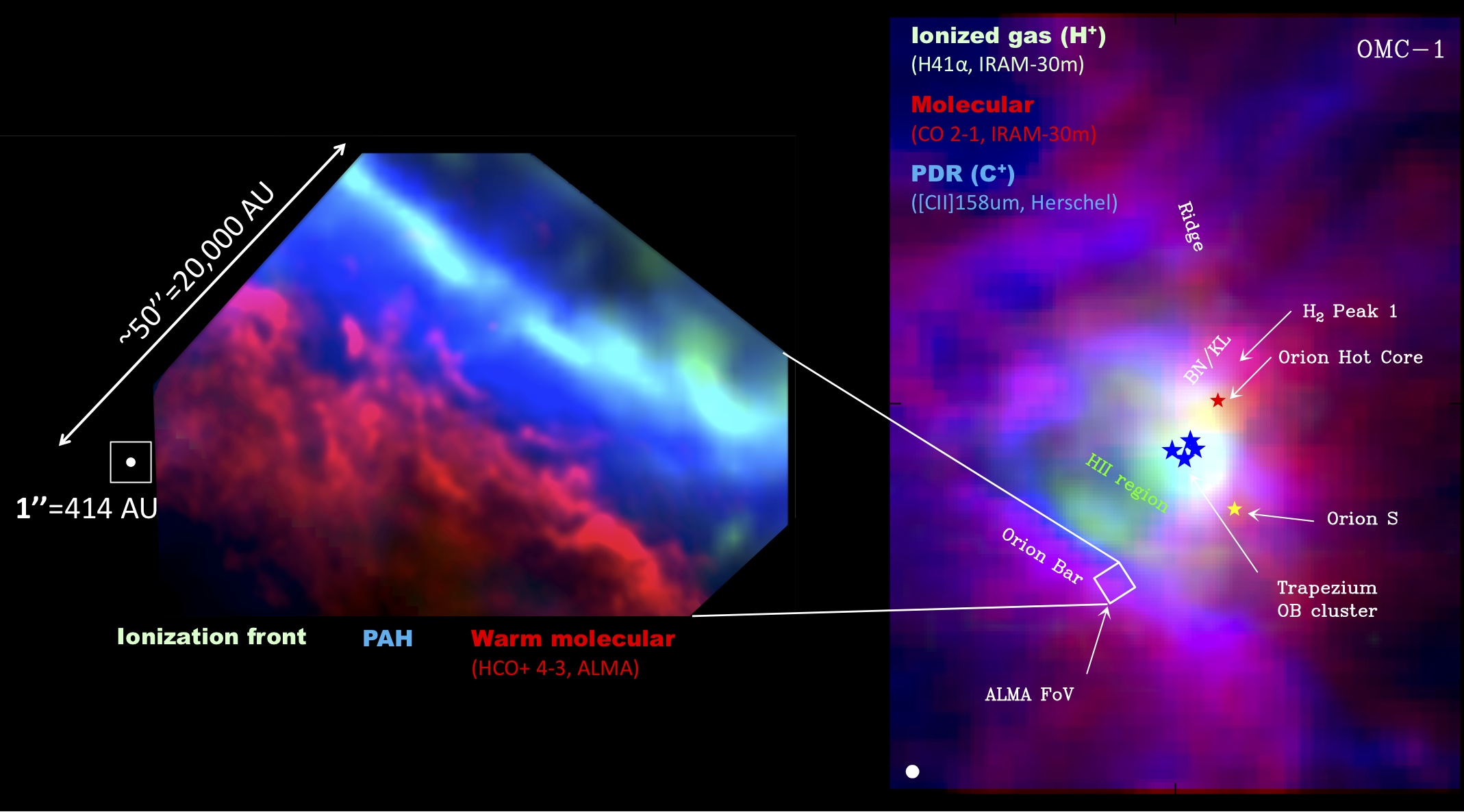}
\vspace{-0.2cm}
\caption{Central parsec ($\sim$\,7.5$'$\,$\times$\,11.5$'$) of the Orion molecular core OMC-1  around the Trapezium  cluster (\textit{right},\,\cite{2015ApJ...812...75G}) and 1$''$ resolution zoom 
\mbox{($\sim$\,50$''$\,$\times$\,50$''$)}  resolving a high pressure 
($P_{\rm th}/k_{\rm B}$\,$\sim$\,10$^8$\,K\,cm$^{-3}$) and nearly edge-on \HII/PDR/molecular cloud 
interface  in the Orion Bar PDR (\textit{left},\,\cite{2016Natur.537..207G}).}
\label{fig-2}       
\end{figure}

\subsection{New-generation observations of the ISM and PDR environments}

 State-of-the-art observations allow us to \mbox{access} a plethora of  multi-line diagnostics  at \mbox{increasingly}  higher sensitivity, angular resolution, and field-of-view.
 Recent \mbox{observations} challenge some of our previous views of PDRs,
 for example, their \mbox{fundamental} small-scale structure(s),
 the origin of the observed rich chemistry,
  and their gas \mbox{dynamics}:  \mbox{propagation} of \mbox{ionization/dissociation} fronts,  \mbox{photoevaporation}, gas compression, and \mbox{radiation} \mbox{pressure} on grains.  
        Last years have seen rapid progress in heterodyne receiver technology and specific  techniques involving (sub)mm spectral-imaging of the ISM:

         \textit{i)} Increased sensitivity and angular resolution of  interferometric mosaics (with ALMA and NOEMA), providing astonishing  sub-arcsecond resolution images of the fundamental structure of the ISM (e.g., see Fig.~2 for the Orion Bar PDR \cite{2016Natur.537..207G}).    
                    
\textit{ii)} Broader instantaneous bandwidth, allowing us to \textit{a)}  map large portions of star-forming clouds in several molecular lines  simultaneously 
(CO, HCN, HNC, HCO$^+$, N$_2$H$^+$, ...\,\cite{2017A&A...599A..98P}), and
\textit{b)} to obtain deep line surveys of the molecular content
(through the detection
of hundreds of molecular lines
\cite{2015A&A...575A..82C,2017A&A...603A.124C})  and physical conditions of representative environments.
 IRAM, its forefront instrumentation, and the variety of available \mbox{observational} techniques (from \mbox{on-the-fly} mapping to line surveys) have played a pivotal role in improving our understanding of the ISM and its underlying PDR processes.
 
\textit{iii)} Development of airborne multi-pixel arrays up to the THz domain, able to map the main gas cooling lines (i.e., the cloud energetics)  at high-spectral resolution and providing, for example, velocity-resolved square-degree maps of the [\CII]158\,$\mu$m emission  \cite{2020A&A...639A...2P}.

As a consequence of these developments, current research on PDRs and associated processes is not only about the detailed study of small fields in bright cloud rims such as the  Orion Bar  -- from which we learn so much -- but more generally about the role of stellar feedback and the evolution of the FUV-illuminated  ISM at all relevant spatial scales: from cores
and giant molecular clouds (GMCs) \cite{2015ApJ...812...75G,2020A&A...639A...2P} to distant star-forming galaxies. \\

\begin{figure}[t]
\centering
\includegraphics[width=12.5cm,clip]{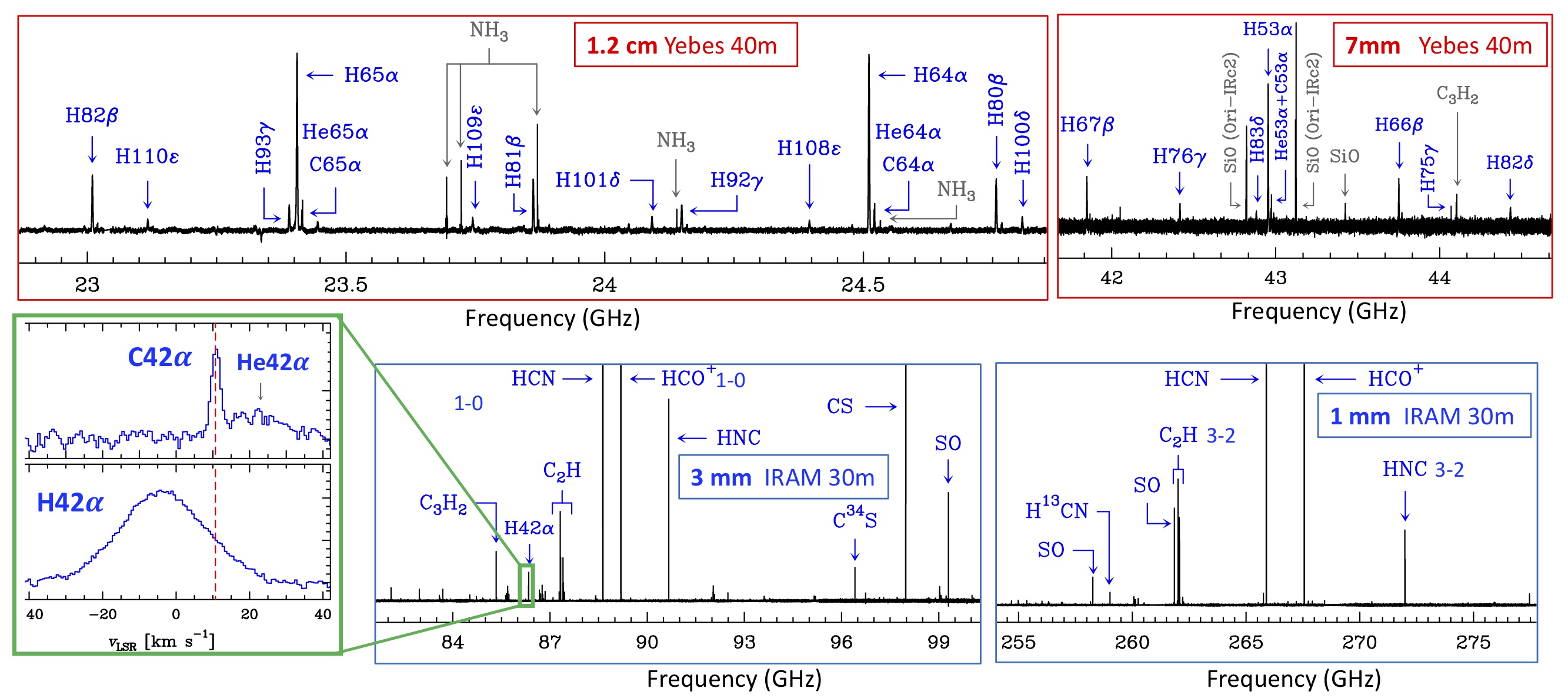}
\caption{Multi-line diagnostics (molecular and recombination lines) 
observed  toward the Orion Bar PDR and adjacent \HII~region
using the IRAM\,30m \cite{2015A&A...575A..82C,2017A&A...603A.124C,2019A&A...625L...3C} and Yebes\,40m \cite{2021A&A...647L...7G} radiotelescopes.}
\label{fig-3}       
\end{figure}

\subsection{Molecular content of PDRs and peculiarities}

PDRs were once thought to be too harsh to develop a rich chemistry. However, our IRAM\,30\,m
and Yebes\,40\,m line surveys  of the Orion Bar, a strongly irradiated PDR \mbox{($G_0$\,$\simeq$\,10$^4$),}
\mbox{exemplify that PDRs} do host multiple lines and unique chemical features
that are not seen (or are too faint to be detected) in regions shielded from FUV radiation. These specific lines \mbox{(Fig.~3)} have a great diagnostic power 
as tracers of FUV radiation from short-lived  massive stars, and thus, of 
star formation: narrow carbon \mbox{radio} recombination lines (with line intensities proportional to 
$n_{\rm e}^{2} T_{\rm e}^{-1.5}$) tracing
the presence of C$^+$ ions and of high ionization fractions 
(\mbox{$n_{\rm e}$/$n_{\rm H}$\, $\approx$\,C$^+$/H}; \cite{2019A&A...625L...3C}),  reactive molecular ions
\mbox{(CH$^+$, SH$^+$, HOC$^+$, and CO$^+$;\,\cite{2013A&A...550A..96N,2017A&A...601L...9G})}, as well as \mbox{enhanced} abundances of \mbox{certain} \mbox{hydrocarbons}  (C$_3$H$^+$, \mbox{$l$-C$_3$H};\,\cite{2015A&A...575A..82C}),  \mbox{radicals} 
(NH \cite{2022arXiv220610441G},  OH \cite{2017A&A...603A.124C}, and HCO \cite{2011A&A...530L..16G}), and
\mbox{unstable} isomers of \mbox{organic} molecules ($cis-$HCOOH;\,\cite{2016A&A...596L...1C}). \mbox{In addition}, specific sulfur-bearing molecules such as H$_2$S show bright emission lines and enhanced abundances
(as revealed by the detection of the rare H$_{2}^{33}$S 
isotopologue;\,\cite{2021A&A...647A..10G}). \mbox{Indeed}, the detection of  sulfur
radio \mbox{recombination} lines \cite{2021A&A...647L...7G}  implies a gas-phase
S$^+$/H abundance
 of \mbox{(1.4\,$\pm$\,0.4)$\cdot$10$^{-5}$}
at edge of the Orion Bar. This is the elemental gas-phase sulfur abundance before an undetermined fraction goes into S-bearing molecules and deplete as grain ice mantles in the cloud 
interior. The inferred sulfur abundance  in the PDR edge matches the solar 
abundance \cite{2021A&A...647A..10G}. It thus leaves little room for 
large  depletions of sulfur  onto rocky grains in molecular clouds.

Some of the above "PDR species" are readily detected in  diffuse clouds
\cite{2019A&A...622A..26G} as well as  in  \mbox{starburst} galaxies \cite{2008A&A...492..675F}. 
These detections prove the interaction of stellar radiation with interstellar matter in different regimes of FUV flux, gas density, and spatial scale.

\section{Models, need of precise atomic and molecular data, and outlook}

The theoretical study and first thermo-chemical models of dense PDRs started to  develop more than 50 years ago. 
PDR models  solve the penetration of FUV radiation into molecular clouds. This 
includes treating the absorption of FUV photons to electronically excited levels of H, H$_2$, and CO, as well as the absorption and anisotropic scattering by dust grains \cite{2007A&A...467....1G}. Among other things, the wavelength-dependent FUV 
($\lambda > 911\AA$) radiation field at different cloud depths determines the rate at which low-ionization-potential atoms are ionized (C, S, Si, Fe, ...) and at which molecules are ionized and dissociated. The FUV field, gas \mbox{density}, and dust/PAH properties set the main heating mechanisms, whereas cooling from  [\CII], [\OI],
H$_2$, and mid-$J$ CO emission lines \cite{2018A&A...615A.129J} determines the cloud-depth variation of the gas temperature. This temperature sets the rates of a large network of gas-phase ion-neutral and neutral-neutral chemical reactions forming and destroying molecules. 
At deeper depths into the molecular cloud, typically \mbox{$A_V$\,$>$\,8\,mag}, the flux of stellar FUV photons is almost completely attenuated and gas-phase molecules and atoms freeze-out on dust grains as temperatures drop. 
Cosmic-rays drive the ionization of atoms and 
molecules, but at much lower rates than in gas directly  exposed to stellar FUV radiation. 
This leads to low \mbox{ionization} fractions and a slower chemistry.  The catalytic formation of molecules on grain surfaces and subsequent thermal  (sublimation) and nonthermal (photodesorption, cosmic-ray induced, or chemical) 
desorption adds a new layer of chemical complexity to model.

\begin{table}
\centering
\caption{Microphysical processes relevant to the study of interstellar PDRs}
\footnotesize
\label{tab-1}       
\begin{tabular}{p{2.5cm} p{3cm} p{2.8cm} p{2.5cm}}
\hline
\textbf{Process} & \textbf{Relevance} &  \textbf{Required parameter} & 
\textbf{Methodology}  \\\hline 
Inelastic\,\,collisions with \mbox{$o$-H$_2$}, \mbox{$p$-H$_2$}, H, and $e^-$ & Non-LTE\,\,excitation. 
\mbox{Precise}\,\,comparison\,\,with observed\,\,line\,\,intensities & 
Inelastic\,\,collisional rate\,\,coefficient 
\mbox{$\gamma_{ij}(T)\sim \sigma_{ij}\,\cdot\,v$\,\,\,(cm$^3$\,s$^{-1}$)} & 
Scattering\,\,calculations.  Laboratory. \\\hline 
 Chemical\,\,reactions  &  
 Formation and destruction of molecules and PAHs &   
 Reaction rate\,\,coefficient 
 \mbox{$k(T)$\,\,\,\,\,(cm$^3$\,s$^{-1}$)} &
   \mbox{Reaction}\,\,dynamics calculations.
  \,\,\,\,\,Laboratory.     \\\hline 
Reactions\,\,with 
\mbox{vibrationally}\,\,excited H$_2$\,($v$$\geq$1,\,$J$) &
Reactions of FUV-pumped 
H$_2$\,\,with\,\,atoms\,\,and 
molecules\,\,overcomes reaction endoergicities &
State\,\,specific\,\,rate coefficient 
 \mbox{$k_{v,\,J}(T)$\,\,\,\,\,(cm$^3$\,s$^{-1}$)}  & 
\mbox{Reaction}\,\,dynamics calculations.
  \,\,\,\,\,Laboratory.  \\\hline 
Photo-ionization and dissociation 
of atoms, molecules, and PAHs & 
Photochemistry\,\,induced by stellar FUV photons &
Photo\,\,cross\,\,sections  
$\sigma_{\rm ion}(\lambda)$,\,\,\,$\sigma_{\rm diss}(\lambda)$\,\,\,(cm$^{-2}$) & 
Experiments
 (e.g.,\,\,synchrotron).  Calculations. \\\hline 
 Low-$T_{\rm e}$\,\,radiative 
 and\,\,dielectronic 
 recombination\,\,of S$^+$\,\,(Si$^+$,\,\,Fe$^+$,\,\,...) &
 Defines S$^+$/S transition.
 Determines  recombination line spectrum &
 Recombination\,\,rate \mbox{coefficients} 
 \mbox{$k_{\rm RR,\,DR}(T_{\rm e})$\,\,\,\,\,(cm$^3$\,s$^{-1}$)} &
 Calculations. \\\hline 
 Adsorption/desorption of\,\,atoms\,\,and\,\,molecules
 onto/from dust grains & 
 Freeze-out,\,\,grain\,\,surface chemistry and
 sublimation of\,\,atoms/molecules\,\,from dust grains &
 Binding\,\,energies 
 $E_b$/$k_B$\,\,\,(K) & 
 Temperature-programmed \mbox{desorption}\,\,experiments.
 Calculations.\\\hline 
 Non-thermal desorption  from grains & 
 Photodesorption\,\,and cosmic-ray\,\,induced desorption of
 ice mantles & 
 Desorption\,\,yields  $Y$\,\,(molecule\,\,photon$^{-1}$)
 $k_{\rm CR}$\,\,(s$^{-1}$) & 
FUV,\,\,\mbox{X-ray},\,\,and \mbox{particle}\,\,irradiation experiments.
  \\\hline 
FUV\,\,extinction

Photoelectric\,\,effect &  
FUV\,\,penetration.  Gas\,\,heating &
 Grain/PAH\,\,optical\,\,props.
 $Q_{\rm abs}$($\lambda$), $Q_{\rm sca}$($\lambda$), $g$($\lambda$) &
Laboratory.
  \\\hline 
\end{tabular}
\end{table}

In order to correctly interpret the amount of information and the many details  provided by state-of-the-art multi-line observations, astrochemical models need to accurately calculate the rates of the above microphysical processes
(interstellar gas is rarely in \mbox{thermal} or \mbox{chemical} equilibrium). This is the main goal of the \textit{Meudon} PDR code
\cite{2006ApJS..164..506L}, a publicly available and open source model 
({\color{blue}{\url{https://ism.obspm.fr/}}}) that is continuously upgraded toward more realistic descriptions of  these processes \cite{2014A&A...569A.100B} and as new rate
coefficients and \mbox{cross sections} are published in the literature.
Hence, it is mandatory to strengthen and foster interdisciplinary collaborations aimed to better characterize these processes, and to determine the precise atomic and molecular data needed in astrochemical models.
These data can be obtained in sophisticated laboratory experiments or through
high-level \textit{ab initio} calculations. \mbox{Table~1} summarizes some of the most relevant microphysical processes taking place in the ISM (with \mbox{emphasis on PDRs}) together with the physical parameters and rate coefficients typically needed in accurate
 thermo-chemical and \mbox{non-LTE} \mbox{excitation}    models.



 PDR models play a central role to correctly interpret spectroscopic observations of a very significant  fraction of the ISM in star-forming galaxies (all neutral
 gas at $A_V$\,$<$\,8\,mag). 
 New instrumentation 
  will allow  mapping 
several square-degree areas of GMCs (reaching the scales that dominate
the extragalactic emission) in  \mbox{multiple} rotational lines, \mbox{critical} to derive accurate physical conditions and precise chemical abundances. More automatic statistical data analysis tools -- designed to link  large grids of astrochemical models and \mbox{hydrodynamic} PDR simulations with observations --
will be used to extract all the information contained in the million spectra these velocity-resolved cubes will generate \cite{2017A&A...599A..98P}. These cubes  provide access to the gas
kinematics, thus to the driving forces in the ISM, and reveal the 
dynamical and non-stationary aspects of stellar feedback in the ISM
 (e.g., \cite{2021A&A...656A..65M}).
 In parallel, \mbox{interferometric} mosaics will allow us to locally zoom into particular fields/templates at very high angular resolution. These images reveal  the spatial scales at which chemical abundances and physical conditions abruptly change in clouds \mbox{(Fig.~2)}. Interferometric observations also spatially resolve GMCs in nearby \mbox{galaxies}, 
with emission features often dominated by gas at low $A_V$.
JWST will soon reveal the evolution of  PAHs, warm dust grains, and H$_2$ emission (e.g., \cite{2022PASP..134e4301B}) in all kind of FUV-irradiated environments. 
The IR to cm emission from these ``PDRs'' is \mbox{determined} by many microphysical processes that form, \mbox{excite}, and destroy the gas and dust in the ISM.  
Knowing the rates of these processes is equally important to correctly understand the many faces of stellar feedback in the ISM.

%
%

\end{document}